# Photoelectrochemical and photocatalytic properties of N + S co-doped TiO$_2$ nanotube array films under visible light irradiation

Guotian Yan, Min Zhang*, Jian Hou, Jianjun Yang*

Key Laboratory for Special Functional Materials of Ministry of Education, Henan University, Kaifeng 475004, PR China

A B S T R A C T

In this paper, we report on the co-doping nitrogen and sulfur has been achieved in the TiO$_2$ nanotube array films by treatment with thiourea and calcination under vacuum at 500 °C for 3 h. The samples were characterized by scanning electron microscopy, X-ray diffraction (XRD), X-ray photoelectron spectroscopy (XPS) and ultraviolet–visible diffuse reflectance spectroscopy. XPS spectra revealed that N might coexist in the forms of N—Ti—O and N—O—Ti, S was incorporated into the lattice of TiO$_2$ through substituting oxygen atoms in the N + S co-doped TiO$_2$ nanotube array films. XRD patterns indicated that improved crystallinity was obtained for N + S co-doped TiO$_2$ nanotube arrays as compared to that of undoped TiO$_2$ nanotube arrays. In photoelectrochemical measurements, the photocurrent of N + S co-doped TiO$_2$ nanotube array films was greatly enhanced compared to that of undoped samples under visible light irradiation. And the photocatalytic activities of the samples were evaluated on the removal of methylene blue under visible light irradiation. The N + S co-doped TiO$_2$ nanotube array films showed a better photocatalytic activity than the undoped sample due to the N, S doping.

## 1. Introduction

TiO$_2$-based photocatalytic materials have been extensively studied over the past three decades due to its low toxicity/cost, chemical stability, excellent photocatalytic activity, and long lifetime of electron/hole pairs. However, one of the main disadvantages of heterogeneous photocatalysis over TiO$_2$ is that its large band gap (about 3.2 eV) can absorb only the ultraviolet (UV) light contained in a solar spectrum and the inability to utilize visible light limits the efficiency of photocatalytic degradation of organic pollutants [1]. Doping TiO$_2$ with nonmetal element (e.g., N, S, C and F) seems to be a feasible way and attracts increasing attention in recent years [2–4]. Nonmetal elements having atomic orbital (e.g., N 2p, S 3p, and C 2p) with a potential energy higher than that of the O2p atomic orbital are introduced into TiO$_2$. In this case, new valence bands can be formed instead of a pure O2p atomic orbital, which results in a decrease in the band gap energy without affecting the conduction band level [5–8]. More recently, the simultaneous doping of two or three kinds of nonmetal atoms into TiO$_2$ has attracted considerable interest since it could result in a higher photocatalytic activity and special physical and chemical characteristics compared with single element doping into TiO$_2$ [9–11].

The photocatalytic activity of TiO$_2$ strongly depends on morphology, crystalline structure, size, and the prepared methods. The anodic TiO$_2$ nanotube array exhibits more promising photoelectrochemical and photocatalytic properties because the nanotube array architecture enhances the electron percolation pathway for vectorial charge transfer, promotes ion diffusion in the semiconductor/electrolyte interface, and restrains photogenerated electron–hole pairs from recombination [12]. Recently, TiO$_2$ nanotube arrays have been investigated in photoelectrochemical and photoelectrocatalytic regions, such as solar cells, photoelectrocatalytic water splitting, and the degradation of organic pollutants [13,14]. However, the utilization efficiency of solar light for TiO$_2$ nanotube arrays is still poor. Some studies have shown that doped TiO$_2$ nanotube arrays with single nonmetals, such as nitrogen [15–17], carbon [18,19], sulfur [7], or boron [12], would increase its visible light photoresponse. The simultaneous doping of carbon and nitrogen into TiO$_2$ nanotube array had also been fabricated [20]. The combination of element doping and TiO$_2$ nanotube structure resulted in a marked increase in the photocatalytic activity of TiO$_2$ under visible light irradiation. To the authors' knowledge, however, no detailed studies on the fabrication and photocatalytic properties of N + S co-doped TiO$_2$ nanotube arrays have been reported. Therefore, in this paper, we report on the preparation of N + S co-doped TiO$_2$ nanotube array films by a facile method and investigation of its photoelectronchemical and photocatalytic properties.

In the present work, TiO$_2$ nanotube arrays were prepared by a two-step electrochemical anodization method. Nitrogen and sulfur co-doping had been achieved in the TiO$_2$ nanotube array films by treatment with thiourea and calcination under vacuum at 500 °C for 3 h. The morphology, structure and optical absorption property of samples were characterized by scanning electron microscopy (SEM), X-ray diffraction (XRD), X-ray photoelectron spectroscopy (XPS) and ultraviolet–visible diffuse reflectance spectroscopy (UV–vis DRS). The effects of N + S co-doping on the photoelectrochemical properties and photocatalytic activity of methylene blue (MB) degradation were investigated for the undoped and N + S co-doped TiO$_2$ nanotube array films under visible light irradiation.

* Corresponding authors.
E-mail addresses: zm1012@henu.edu.cn (M. Zhang),
yangjianjun@henu.edu.cn (J. Yang).

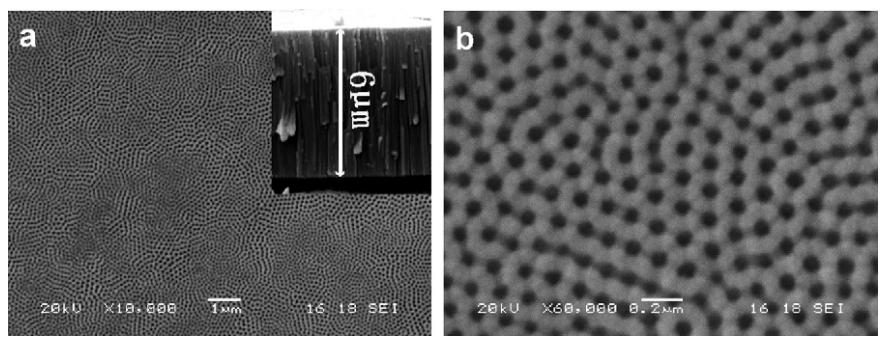

**Fig. 1.** SEM top view image and cross-section (inset picture) of N + S co-doped TiO$_2$ nanotube array: (a) low magnification, (b) high-magnification.

## 2. Experimental

### 2.1. Fabrication of N + S co-doped TiO$_2$ nanotube arrays

Titanium plate (purity >99.6%) was purchased from Baoji Boxin Metal Materials Ltd. Company, China. Titanium sheets (20 mm × 40 mm of 0.25 mm thick) were firstly sonicated in acetone, isopropanol and methanol for 10 min in sequence before experiments. Then the titanium sheets were etched in a mixture of HF/HNO$_3$/H$_2$O (1:4:5 in volume) for 20 s, rinsed with deionized water and dried under a N$_2$ stream.

The self-organized and well-aligned TiO$_2$ nanotube arrays were fabricated by two-step electrochemical anodization process. A rectangle Ti sheet was used as an anode and Pt meshwork as a cathode in the anodic oxidation experimental set-up. A direct current power supply is utilized for the control of experimental voltage and a mixed electrolyte solution of ethylene glycol containing 0.25 wt% NH$_4$F and 2 vol% deionized water as electrolyte in the electrochemical process. The first oxidation step was conducted at 60 V for 1 h, and removed the surface membrane shedding by sonication in distilled water, and then dried under high purity N$_2$ stream in room temperature to use for next step. The second oxidation step: the Ti substrates obtained after the first step treatments were oxidized in the original electrolyte at 60 V for 0.5 h. After anodization, the samples were sonicated (removing the surface cover) in ethanol for 8 min, and dried under high purity N$_2$ stream to obtain the as-prepared TiO$_2$ nanotube arrays. A circulation pump maintained a low temperature of about 20 °C throughout the experiment and the oxidation process was always accompanied by magnetic stirring.

N + S co-doped TiO$_2$ nanotube arrays were prepared by treating as-prepared TiO$_2$ nanotube arrays with thiourea under heating. As-prepared TiO$_2$ nanotube array films and thiourea was placed in a tube furnace (OTF-1200X), and then the furnace was pumped by an oil rotary vane type pump. When the vacuum level reached 0.1 MPa in the furnace, the temperature of the furnace increased to 500 °C with heating rate of 10 °C min$^{-1}$, held for 3 h and cooled naturally to room temperature.

For comparison, the as-prepared TiO$_2$ nanotube array films were calcinated in air at the same temperature to yield undoped TiO$_2$ nanotube array samples.

### 2.2. Characterization

The morphologies of TiO$_2$ nanotube array films were characterized with a SEM, (JSM5600LV, Japan). A scanning UV–vis spectrophotometer (Cary 5000, Varian, America) equipped with a Labsphere diffuse reflectance accessory was used to collect the adsorption spectra of undoped and N + S co-doped TiO$_2$ nanotube array films in the range of 300–700 nm at a scan speed of 300 nm min$^{-1}$. Chemical state analysis was carried out using Axis Ultra XPS (Kratos, UK). A monochromatic Al source was utilized operating at 210 W with a pass energy of 20 eV and a step of 0.1 eV. All XPS spectra were corrected using the C 1s line at 284.6 eV. XRD experiments were carried out using a Philips X-ray diffractometer (X'Pert Pro PW3040/60, Holland) to determine the structure of the samples.

### 2.3. Photocurrent and photocatalytic activity measurements

The photoelectrochemical properties were characterized using a three-electrode photoelectrochemical cell with undoped or doped TiO$_2$ nanotube array films as the working electrode, an Ag/AgCl electrode as the reference, and a platinum meshwork as the counter electrode in 1 mol L$^{-1}$ KOH solution. The samples were pressed against a semi-cylindrical in the photoelectrochemical cell, leaving an area of 4 cm$^2$ exposed to the light source through a quartz window. A 300 W Xenon lamp (PLS-SEX300, Beijing Changtuo) was used as the light source; a cut off filter was used to remove any radiation below 420 nm to ensure irradiation by visible light only. A scanning potentiostat (IM 6ex, Germany) was used to perform a potentiodynamic scan from −0.4 to 0.3 V vs SCE at a rate of 10 mV S$^{-1}$ and measure the generated current. The photocurrent transient was also measured at a fixed bias potential, 0.2 V vs SCE, with a light pulse of 50 s under visible light irradiation.

The photocatalytic activities of undoped and N + S co-doped TiO$_2$ nanotube array film were evaluated by analyzing photocatalytic degradation of MB in aqueous solution. Visible light with vertical irradiation was used as light source. Undoped or N + S co-doped TiO$_2$ nanotube array film was dipped into 50 mL of MB solution with an initial concentration of 10 mg L$^{-1}$. The effective area of TiO$_2$ nanotube array films for photocatalytic reaction was 4 cm$^2$. Prior to photoreaction, the solution was kept for 30 min in the dark to reach adsorption–desorption equilibrium. During the photocatalytic reaction, the absorbance of MB was measured at a time interval of every 20 min using a sp-2000 spectrophotometer at 664 nm.

## 3. Results and discussions

### 3.1. SEM analyses

Fig. 1 shows low- and high-magnification SEM images (top view and a cross-section as inset) of the TiO$_2$ nanotube arrays after the thiourea treatment. It can be seen that the nanotubes of N + S co-doped TiO$_2$ nanotube arrays are open at the top with an average diameter of ca. 100 nm, which is the same as that for an undoped TiO$_2$ nanotube arrays. No distinct deposit appeared on the surface of TiO$_2$ nanotube arrays films. The cross-sectional image of the N + S co-doped sample indicates that the self-organized and ordered nanotube arrays are perpendicular to the substrate with a thickness of ca. 6.0 μm.

### 3.2. XRD patterns analyses

The XRD patterns of as-prepared, undoped and N + S co-doped TiO$_2$ nanotube array films are presented in Fig. 2. The results indicate that as-prepared TiO$_2$ nanotube arrays are amorphous. Undoped and N + S co-doped TiO$_2$ films gave rise to well-established peaks of the anatase phase (JCPDS card No. 84-1286), which can be ascribed to the (1 0 1), (0 0 4), (2 0 0), (1 0 5), (2 1 1), (2 0 4), (1 1 6), (2 2 0), (2 1 5), (2 2 4) planes of anatase TiO$_2$ (Fig. 2b and c). Both annealing and doping treatments made TiO$_2$ nanotube array films transform from an amorphous structure to the anatase phase. The (1 0 1) anatase peak intensity of the N + S co-doped sample is greatly larger than that of undoped sample, which demonstrates that the crystallinity of TiO$_2$ could be improved by N + S co-doping. The particle sizes of undoped and N + S co-doped samples are 15.5 nm and 17 nm calculated by Scherrer formula,

### 3.3. XPS analyses

High-resolution XPS of N 1s, S 2p, Ti 2p and O 1s core levels of the N + S co-doped $TiO_2$ nanotube array films were measured to obtain detailed chemical state information. Fig. 3a shows the XPS spectra for the N 1s region of N + S co-doped $TiO_2$ nanotube array films and its fitting curves. It is a broad peak from 396 eV to 403 eV. Deconvolution of the N 1s peak revealed the presence of two peaks with binding energy at 398.4 and 399.9 eV. The N 1s peak localized at 398.4 eV can be ascribed to anionic N incorporated in $TiO_2$ in O—Ti—N linkages, while the peak at 399.9 eV can be attributed to the presence of oxidized nitrogen such as N—O—Ti from the previous reports [1,5,6,22]. Therefore, the XPS results indicated that two forms of O—Ti—N and N—O—Ti coexist in doped $TiO_2$ nanotube array films.

Fig. 3b shows the high-resolution XPS spectra of the S 2p region for the doped $TiO_2$ nanotube array films. The sulfur atoms are in the $S^{2-}$ state, with a peak at about 163.8 eV corresponding to the anionic $S^{2-}$ in Ti—S bond formed when some of the oxygen atoms in the $TiO_2$ lattice are replaced by sulfur atoms [23]. The peaks around 168.4 eV can be attributed to $S^{6+}$. This peak is often assigned to the presence of $SO_4^{2-}$ ions produced during the combustion of thiourea on the titania surface in the recent report [9]. The XPS results confirmed the successful substitution of oxygen by sulfur in $TiO_2$ nanotube array films. Theoretical calculations already demonstrated that this substitution would result in a band gap narrowing of $TiO_2$ samples [24].

Fig. 3c and d present high-resolution spectra of Ti 2p and O 1s of undoped and N + S co-doped $TiO_2$ nanotube array films. Compared with the undoped $TiO_2$ nanotube array films, the Ti 2p and O 1s peaks of co-doped $TiO_2$ samples are slightly shifted toward lower binding energy. It has been reported that the incorporation of N atom into the $TiO_2$ lattice can also lead to the shift of Ti $2p_{3/2}$ respectively. These results show that N + S co-doped samples have the larger average crystalline size and the higher anatase crystallinity. Correspondingly, the grain boundaries and amorphous regions that can serve as charge-carrier recombination centers are reduced, which may favor an increase of photocurrent and photocatalytic activity [21]. Moreover, the (1 0 1) plane of anatase peaks of the N + S co-doped sample slightly shift to higher values of $2\theta$, in comparison with those of its annealed counter part. This shift suggests that the oxygen or Ti atoms in the lattice of anatase in doped samples may be substituted by N and S atoms. This result is in agreement with the work reported by Lee and co-workers [11].

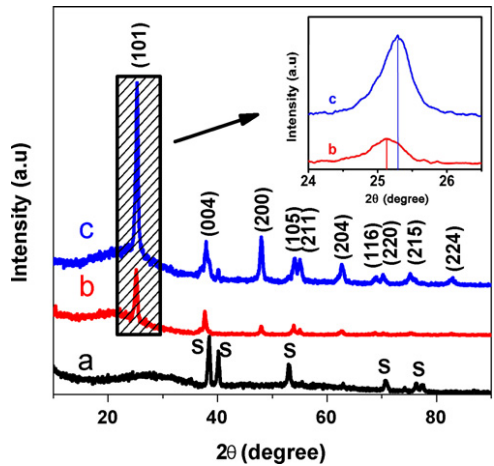

**Fig. 2.** XRD patterns of samples: (a) as-prepared $TiO_2$ nanotube array, (b) undoped $TiO_2$ nanotube array, (c) N + S co-doped $TiO_2$ nanotube array; S: Ti substrate.

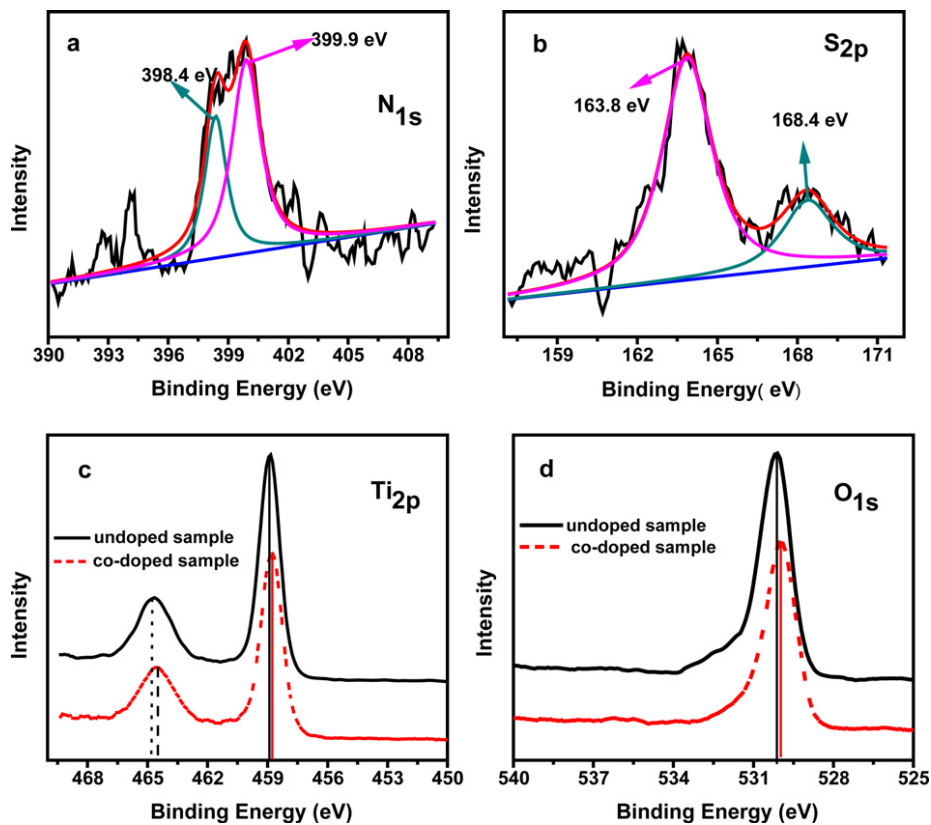

**Fig. 3.** High-resolution XPS spectra of N 1s (a) and S 2p (b) of N + S co-doped $TiO_2$ nanotube array and high-resolution XPS spectra of Ti 2p (c) and O 1s (d) of undoped and N + S co-doped $TiO_2$ nanotube array.

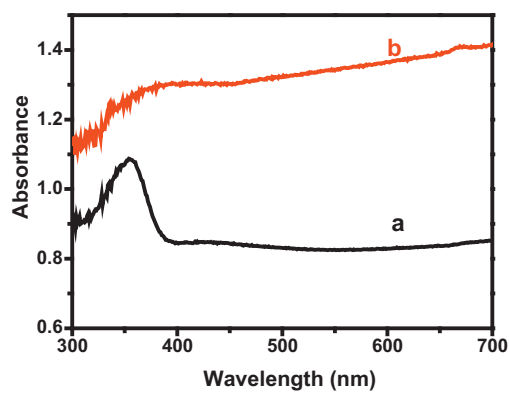

**Fig. 4.** UV–vis DRS spectra of undoped $TiO_2$ nanotube array (a) and N + S co-doped $TiO_2$ nanotube array (b).

to lower binding energy [25]. The different electronic interactions of Ti with N anions by co-doping causes partial electron transformation from the N to Ti and an increase of the electron density on Ti because of the lower electro negativity of nitrogen compared with oxygen [26,27]. On the other hand, the shifts of Ti $2p_{3/2}$ and the O 1s peaks are due to the introduction of oxygen vacancies into the $TiO_2$ lattice [28,29]. In our related experiments, the formation of single-electron-trapped oxygen vacancy in the structure of [O—Ti—O—Ti—] have been confirmed for N-doped $TiO_2$ sample using electron spin resonance (ESR) [30]. $TiO_2$ can be partially reduced under high temperature and vacuum, resulting in the shifts of both the Ti $2p_{3/2}$ and the O 1s peaks to lower binding energies [6]. These XPS results indicate that the N + S co-doping is successfully obtained and accompanied by oxygen vacancy formation. This substitution was also confirmed by XRD as the peaks of N + S co-doped $TiO_2$ nanotube array films shifts slightly to higher $2\theta$ values compared to undoped $TiO_2$ nanotube array films as shown in Fig. 2.

### 3.4. UV–vis DRS spectra analysis

Fig. 4 shows UV–vis DRS spectra recorded from undoped (curve a) and N + S co-doped $TiO_2$ nanotube arrays (curve b). The results indicate the adsorption edge of undoped $TiO_2$ nanotube arrays is around 390 nm, and co-doping nitrogen and sulfur extended the absorption of $TiO_2$ nanotube array films into the whole visible light range. Compared with the undoped $TiO_2$ nanotube array, the anatase absorption becomes less predominant, and the visible light absorption ability increases in co-doped samples. Various mechanisms have been proposed to explain the visible light absorption of N-doped $TiO_2$ [13,20,30–33]. The corporate contributions of nitrogen doping and oxygen vacancies on the optical absorption spectra of $TiO_2$ were confirmed for N-doped $TiO_2$ sample according to spin-polarized density functional-theory calculations by Payne and co-workers [32]. Moreover, the decomposition of thiourea in the heat treatment under vacuum can result in the deposit of carbon on the surface of $TiO_2$ nanotube arrays, which also provided significant absorption in the visible light region [20,33]. Therefore, the remarkable visible light adsorption of the N + S co-doped $TiO_2$ sample is ascribed to the synergetic effects of substitution of the crystal lattice oxygen by nitrogen and sulfur, existence of oxygen vacancies and adsorption of the carbon on the surface $TiO_2$ nanotube arrays film.

### 3.5. Photoelectrochemical properties

It is well known that the interband electron transition is accompanied by relaxation and recombination in $TiO_2$ semiconducting material, and consequently, only part of the photons absorbed by

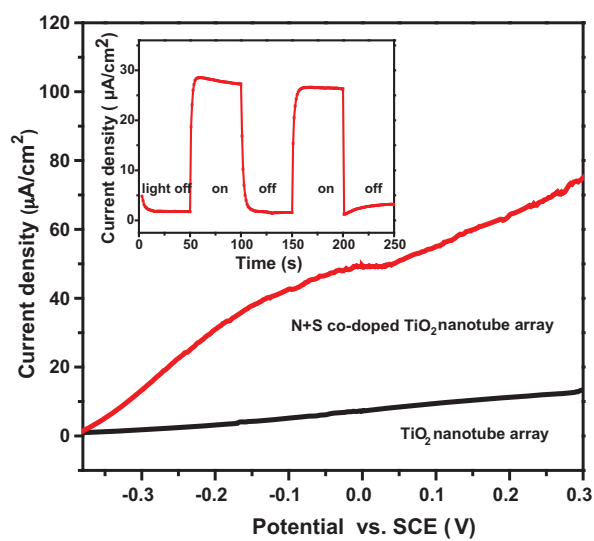

**Fig. 5.** Dependence of the photocurrent on the potential applied for the undoped and the N + S co-doped $TiO_2$ nanotube array under visible light irradiation in a 1 M KOH solution. The inset shows the transient photocurrent generated under pulse visible light irradiation at a fixed bias potential of 0.2 V (SCE) for the N + S co-doped $TiO_2$ nanotube array.

$TiO_2$ can contribute to the generation of a photocurrent [12]. Therefore, it is necessary to take into account the photocurrent–voltage characteristics for further investigation of the photoelectrochemical properties of N + S-doped $TiO_2$ nanotube array films. Fig. 5 shows the photocurrent-potential dependence of the undoped and N + S co-doped $TiO_2$ nanotube arrays under visible light irradiation. The photocurrent density of N + S co-doped $TiO_2$ nanotube arrays photoelectrode is significantly higher than that of an undoped $TiO_2$ nanotube array photoelectrode at all bias potential, indicating a lower recombination of photogenerated electrons and holes after N + S co-doping. The generated photocurrent also increased with increasing the applied bias potential. The higher the applied potential, the higher the photocurrent generated by the N + S co-doped $TiO_2$ nanotube arrays photoelectrode. For the annealed sample without doping, no significant photocurrent was measured over the entire voltage range because of the energy of the irradiation light is smaller than the band gap energy of anatase $TiO_2$. The small observed photocurrent of the undoped nanotube arrays could be ascribed to possible pollutants on electrode surfaces or impurities or species in the KOH solution [7]. The high photocurrent conversion efficiency of N + S co-doped $TiO_2$ nanotube array films may be attributed to improved optical absorption and increased in the crystallinity caused by the co-doping [12].

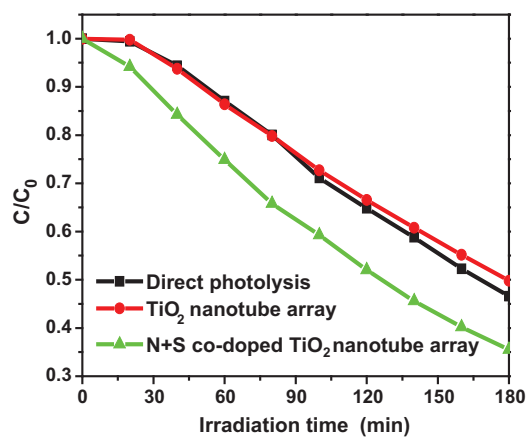

**Fig. 6.** Process of photocatalytic degradation of methylene blue (MB) under visible light irradiation.

Moreover, the transient photocurrent measured at a fixed bias potential of 0.2 V vs SCE with a visible light pulse of 50 s is also shown in the inset of Fig. 5. When visible light reaches the N + S-doped $TiO_2$ nanotube arrays, the transient current quickly increases to 28 $\mu A\,cm^{-2}$ and remains constant until the irradiation is turned off, when the current decays to the background current density (2 $\mu A\,cm^{-2}$). The results indicate that the N + S co-doped $TiO_2$ nanotube arrays are sensitive to visible light and can generate a sustainable steady photocurrent under visible light irradiation.

Higher photocurrent means more photoinduced electrons transferred from N + S co-doped $TiO_2$ nanotube arrays to the counter electrode. The photocatalytic activity of $TiO_2$ usually depends on a competition between the transfer rate of surface charge carriers from the interior to the surface and the recombination rate of photogenerated electrons and holes, a faster photogenerated electrons and holes transfer leading to a higher photocatalytic activity of $TiO_2$.

### 3.6. Photocatalytic activity

The photocatalytic activities of the samples were evaluated by measuring the photocatalytic degradation of MB aqueous solution under visible light irradiation (Fig. 6). The $C/C_0$ vs irradiation time curves showed noticeable differences. 53% of MB was removed by direct photolysis in the absence of $TiO_2$ nanotube array films after 3 h. Under the same conditions, the N + S co-doped sample can degrade MB over 66%, while only 51% of the MB removal is obtained using undoped $TiO_2$ sample. The same removal efficiencies of direct photolysis and photocatalytic degradation of undoped $TiO_2$ nanotube array indicates there is no photocatalytic activity for the undoped sample under visible light irradiation due to the lower energy of visible light relative to the band gap energy of anatase $TiO_2$. For N + S co-doped $TiO_2$ nanotube array film, the degradation ratio can be remarkably increased compared to that of direct photolysis and undoped sample. The higher photocatalytic activity in the N + S co-doped $TiO_2$ nanotube array films can be mainly attributed to the increased visible light absorption and the improved separation of photogenerated electrons and holes by N and S co-doping. Moreover, the surface defects induced by the incorporated N and S atoms are able to serve as catalytic centers which could enhance the photocatalytic activity of N + S co-doped $TiO_2$ nanotube array films.

### 4. Conclusions

In this paper, $TiO_2$ nanotube array films were doped with nitrogen and sulfur using a non-destructive thermal treatment with thiourea at 500 °C under vacuum. XRD spectra indicated both annealing and doping treatments made $TiO_2$ nanotube array films transform from an amorphous structure to the anatase phase. The crystallinity of $TiO_2$ nanotube array can be remarkably improved by N + S co-doping. XPS analysis revealed that N might coexist in the forms of N—Ti—O and N—O—Ti, S was incorporated into the lattice of $TiO_2$ through substituting oxygen atoms in the N + S co-doped $TiO_2$ nanotube array films. UV–vis DRS spectra showed co-doping nitrogen and sulfur effectively extended the absorption of $TiO_2$ nanotube array films into the whole visible light range. Under visible light irradiation, the N + S co-doped $TiO_2$ nanotube array films showed higher photocurrent and photocatalytic activity than undoped $TiO_2$ nanotube array films. These improved properties can be attributed to the increased visible light absorption, the improved photogenerated electrons–hole separation and the surface defects produced by N and S co-doping.

### Acknowledgements

The authors thank the experimental help from Yunge Li, the National Natural Science Foundation of China (no. 20973054) and the Natural Science Project of Science and Technology Department of Henan Province (no. 112300410171) for financial support.